\documentclass[prb,floatfix,twocolumn,superscriptaddress,showpacs,amsmath,amssymb,showpacs]{revtex4-2}

\usepackage{graphicx}
\usepackage{natbib}
\usepackage{amsfonts,amssymb,amsmath}
\usepackage{xcolor}
\usepackage{bm}
\usepackage{bbm}
\usepackage{multirow}
\usepackage{hyperref}

\DeclareMathOperator{\Tr}{Tr}

\begin{document}
\title{Quantics tensor cross interpolation for high-order strong-coupling expansions}
\author{Kanghyeon Kim}
\affiliation{Department of Physics and Chemistry, DGIST, Daegu 42988, Republic of Korea}
\author{Lei Geng}
\email{lei.geng@unifr.ch}
\affiliation{Department of Physics, University of Fribourg, 1700 Fribourg, Switzerland}
\author{Philipp Werner}
\email{philipp.werner@unifr.ch}
\affiliation{Department of Physics, University of Fribourg, 1700 Fribourg, Switzerland}
\author{Aaram J. Kim}
\email{aaram@dgist.ac.kr}
\affiliation{Department of Physics and Chemistry, DGIST, Daegu 42988, Republic of Korea}

\begin{abstract}
Real-time impurity solvers enable the study of transport phenomena and the description of nonequilibrium lattice systems within the framework of dynamical mean-field theory (DMFT). They also provide direct access to the spectral functions of both equilibrium and nonequilibrium systems. A widely used approach is the self-consistent strong-coupling expansion, whose lowest-order implementation corresponds to the non-crossing approximation. Higher-order implementations, however, are computationally demanding because the number of diagram topologies grows factorially with expansion order, while the evaluation of self-energies and Green's functions requires increasingly high-dimensional integrations. Here, we demonstrate that the latter challenge can be mitigated by employing quantics tensor cross interpolation in a variable-separated framework. Compared with the previously used scale-separated approach, the new scheme yields substantially lower bond dimensions and capacitates self-consistent steady-state DMFT calculations up to fourth order. We illustrate its performance with representative results for both equilibrium and photo-doped systems. In addition, we analyze the convergence of the strong-coupling expansion in the challenging noninteracting limit by computing diagrams up to sixth order. At this order, the onset of the asymptotic regime of the strong-coupling expansion becomes apparent, which allows the application of extrapolation techniques.
\end{abstract}

\maketitle

\section{Introduction}
Simulations of the real-time dynamics of strongly correlated electron systems provide direct access to real-frequency spectral functions, thereby circumventing the notoriously difficult analytic continuation of Matsubara axis data \cite{Jarrell1996}. At the same time, they offer a natural framework for investigating nonequilibrium phenomena.
Accurate and efficient real-time computational methods are essential for connecting microscopic model calculations to intriguing experimental observations, such as superconducting-like states in laser-driven systems~\cite{Fausti2011,Mitrano2016,Buzzi2020} or bad metal transport properties observed in cold-atom experiments~\cite{Brown2019}.
For example, on the basis of Hubbard model calculations, $\eta$-pairing \cite{Yang1989} was proposed as a possible mechanism for laser-driven transient superconductivity \cite{Kaneko2019,Li2020}, and shown to successfully reproduce characteristic signatures in the optical conductivity~\cite{Li2020,Li2021b,Geng2026,Werner2026}.

Among the various theoretical methods, dynamical mean-field theory (DMFT)~\cite{Georges1996,Aoki2014} is a promising approach for the accurate description of high-dimensional non-equilibrium systems.
However, the lack of efficient non-equilibrium impurity solvers remains a major bottleneck for its widespread application.
Despite significant progress in the development of numerically exact nonequilibrium impurity solvers, based on continuous-time \cite{Werner2009} and inchworm~\cite{Cohen2015,Erpenbeck2023} Monte Carlo, as well as the influential functional (IF) approach~\cite{Thoenniss2023,Guo2024,Nayak2025}, the implementation of computationally efficient and accurate methods, and their integration into the nonequilibrium DMFT framework, remains an active research area.
Diagrammatic methods have recently profited from efficiency gains related to tensor cross interpolation (TCI)~\cite{Oseledets2010} and its quantics version (QTCI)~\cite{Shinaoka2023,Ritter2024}. These techniques enable a low-rank factorization of Feynman diagrams, and hence the calculation of integrals over the diagram coordinates at a cost that scales linearly with diagram order. 
This approach was first applied to the weak-coupling expansion~\cite{Fernandez2022,Jeannin2025,Matsuura2026} and subsequently extended to its strong-coupling counterpart~\cite{Eckstein2024,Kim2025,Geng2025}.

The strong-coupling diagrammatic expansion of quantum impurity models~\cite{Keiter1971,Pruschke1989} has several appealing features. First, by expanding with respect to the impurity-bath hybridization and resumming diagrams into renormalized impurity propagators, one can achieve rapid convergence of the diagrammatic series for strongly correlated systems and in particular Mott insulators.  
Second, at low orders, the expansion provides a numerically efficient route to nonequilibrium dynamics \cite{Eckstein2010nca}. 
Third, upon convergence at a given order, the strong-coupling approximation is $\Phi$-derivable and therefore conserving in the Baym–Kadanoff sense~\cite{Baym1961}.

However, for metallic solutions and symmetry-broken phases the convergence properties of the strong-coupling approach have not been systematically investigated, mainly due to the limited diagram order, which in DMFT studies has remained $\lesssim 3$~\cite{Eckstein2010nca,Eckstein2024,Kim2025,Geng2025}.
In this work, by optimizing the function parametrization in QTCI, we reach diagram orders up to six and begin to observe the asymptotic behavior of the series expansion, enabling controlled extrapolations to the infinite-order limit in wide parameter regimes.

The paper is structured as follows. 
In Sec.~\ref{sec:methods}, we formulate the diagram construction of the general $X$-order contribution, describe the computational techniques employed in our implementation, and demonstrate the improved performance of the so-called variable-separated parametrization.
In Sec.~\ref{sec:results}, we benchmark for the $U=0$ limit by introducing an extrapolation procedure to the infinite-order limit.
We also apply the high-order impurity solvers to equilibrium and photo-doped states with and without spontaneous symmetry breaking.
Section~\ref{sec:conclusions} contains the conclusions. 

\section{Formalism and implementation}\label{sec:methods}
\subsection{General \texorpdfstring{$X$}{X}-order approximation (\texorpdfstring{$X$}{X}OA)}
\begin{figure}[t]
	\centering
	\includegraphics[width=0.5\textwidth]{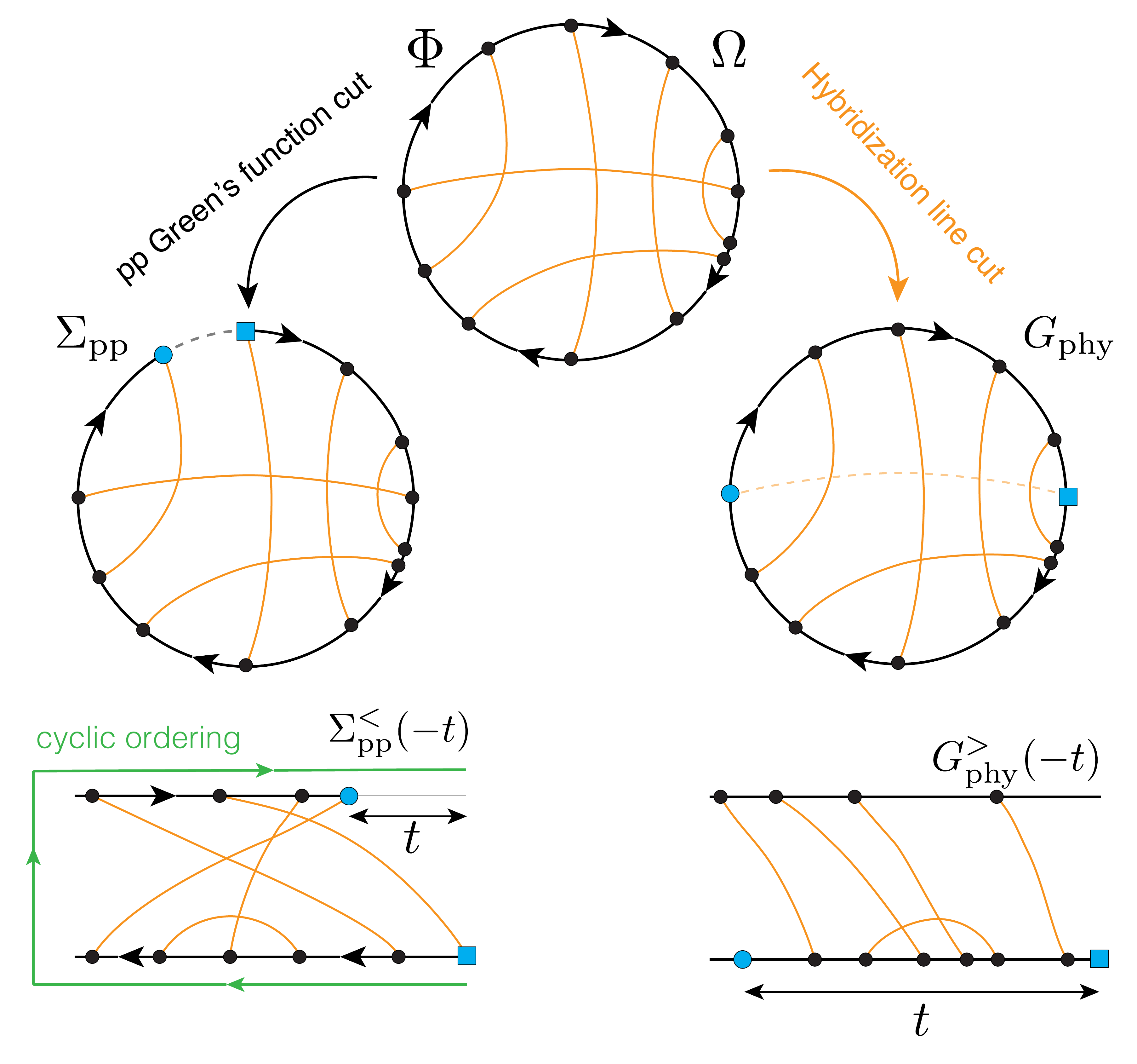}
	\caption{
		Diagram generation from the Luttinger-Ward $\Phi$ or the grand potential $\Omega$. The PP self-energy (physical Green's function) diagram is obtained by cutting a PP Green's function (hybridization function) line. 
		Blue squares (circles) represent impurity creation (annihilation) operators. 
		The bottom panels show examples of $\Sigma_{\text{pp}}$ and $G_{\text{phy}}$ contributions represented on the Keldysh contour. Internal vertices (black dots) are integrated over the contour, respecting the cyclic ordering.
	}
	\label{fig:diagram}
\end{figure}
In this section, we summarize the general $X$-order approximation of the strong-coupling expansion of the Anderson impurity model~\cite{Keiter1971,Pruschke1989, Eckstein2010nca}.
Starting from the Luttinger-Ward functional $\Phi$ or the grand potential $\Omega$ composed of renormalized pseudo-particle (PP)~\cite{Barnes1976} Green's functions $\mathcal{G}$ and hybridization functions $\Delta$, one can deduce the diagrammatic expressions for the PP self-energies and the physical Green's functions by cutting a PP line or a hybridization line, respectively.
Figure~\ref{fig:diagram} presents a schematic picture of the corresponding diagram generation.

Cutting a $\mathcal{G}$ line out of the closed diagrams of the order-$X$ functional $\Phi^{(X)}[\mathcal{G},\Delta]$, equivalent to taking a functional derivative with respect to $\mathcal{G}$, yields a PP self-energy diagram $\Sigma_{\text{pp}}^{(X)}$.
The remaining \textit{backbone} of PP Green's functions can be formally defined by the sequence $\mathcal{B}=\left\{\bm{F}^{\alpha_i}\right\}$ of $2X$ fermionic operators and the $2X$ contour time points $\mathcal{Z}=\left\{z_i\right\}$. Here, $\alpha_i$ is a combined index which includes the spin and creation/annihilation type of the $i$th fermionic operator. The sequence of backbone operators follows the cyclic ordering [Fig. ~\ref{fig:diagram}] along the Keldysh contour, which, starting from the zero time point of the lower branch and passing through the fictitious joint between the lower and upper branches at $-\infty$ (practically set to $-t_{\text{max}}$) ends at the zero time point on the upper branch.
Independent from the backbone, we define the diagram topology by a set of index pairs of fermionic operators $\mathcal{T}=\left\{(a_k,b_k)\right\}$, where the $k$th pair represents a hybridization line connecting the $a_k$th and $b_k$th fermionic operators of the backbone.
The topology $\mathcal{T}$ should satisfy the \textit{two-particle irreducibility} (2PI) in order to avoid double counting of PP self-energy corrections.
By construction~\cite{Luttinger1960}, the 2PI condition for both the PP self-energy and physical Green's function diagrams is satisfied if we cut a single (PP or hybridization) line of a 2PI Luttinger-Ward functional (or grand potential). 
Note that our definition of topology leaves the direction of the hybridization function to be determined by the operator types in the backbone configuration.
In the normal state, the hybridization function can be non-zero only when the selected pair is composed of one creation and one annihilation operator. 
The hybridization direction is then automatically determined as pointing from the annihilation operator to the creation operator.
In a superconducting state, e.g. a spin-singlet $\eta$-pairing state, the anomalous component of the hybridization function is nonzero for pairs of either creation or annihilation operators.

Each diagram defined by the backbone $\mathcal{B}$, the topology $\mathcal{T}$, and the contour time sequence $\mathcal{Z}$ contributes to the final PP self-energy of order $X$ as 
\begin{equation}
	\bm{\Sigma}_{\text{pp}}^{(X)}(z_{2X-1},z_0) = \sum^{}_{\mathcal{B},\mathcal{T}}\int_{}^{}\prod_{i=1}^{2X-2}dz_i~\bm{\sigma}^{(X)}_{\mathcal{B},\mathcal{T}}(\mathcal{Z})~,
	 \label{eq:sigma_pp}
\end{equation}
where
\begin{align}
	\bm{\sigma}^{(X)}_{\mathcal{B},\mathcal{T}}(\mathcal{Z})= \,\,&
	i^X (-1)^{N_\times(\mathcal{T})+N_<(\mathcal{B},\mathcal{T},\mathcal{Z})} \nonumber\\
	&\times \left[ \prod_{k=0}^{X-1} \Delta(z_{a_k},\alpha_{a_k}; z_{b_k},\alpha_{b_k}) \right] \nonumber\\
	&\times \left[\bm{F}^{\alpha_{2X-1}} \prod_{i=0}^{2X-2}\bm{\mathcal{G}}(z_{i+1}, z_{i}) \bm{F}^{\alpha_i}\right].
	\label{eqn:underlyingSigma_pp}
\end{align}

$\Delta(z_i,\alpha_i; z_j,\alpha_j)$, which can be zero, represents the hybridization function determined by the end points and operator types and $\bm{\mathcal{G}}(z_i,z_j)$ the PP propagator starting from contour time point $z_j$ and ending at $z_i$.
$N_\times(\mathcal{T})$ and $N_<(\mathcal{B},\mathcal{T},\mathcal{Z})$ denote the number of crossings between hybridization lines and the number of hybridization lines having opposite direction compared to the backbone direction. 
The appropriate component (lesser or greater) of each hybridization function $\Delta$ and PP propagator $\bm{\mathcal{G}}$ is determined by the contour-branch indices and, when both time arguments lie on the same branch, by their real-time ordering~\cite{Kim2025}.
Note that $\bm{\sigma}$, $\bm{F}$, and $\bm{\mathcal{G}}$ are matrices expressed in some basis of the local Hilbert space, e.~g., $\{| m \rangle = |0\rangle, |\!\!\uparrow\rangle, |\!\!\downarrow\rangle, |d\rangle\}$ (empty state, singly occupied states with spin up and down, and doubly occupied state) for the single-orbital impurity problem. All matrix products should follow the ordering of the corresponding vertices along the Keldysh contour.

The physical Green's function can be obtained in a similar manner, but in this case, we cut a hybridization line instead of a backbone segment from the closed diagrams of the grand potential.
As a result, the backbone of the grand potential retains its closed loop, mathematically equivalent to the trace over the impurity Hilbert space, while the topology $\mathcal{T}$ is replaced by $\mathcal{T}'_l$, which excludes the $l$th pair of $\mathcal{T}$.
The fermion operators originally attached to the missing hybridization function determine the components and the time arguments of the resulting physical Green's function.
The expression for the $X$-order physical Green's function becomes 
\begin{equation}
	G^{(X)}(z_{a_l},\alpha_{a_l};z_{b_l},\alpha_{b_l}) = \sum^{}_{\mathcal{B},\mathcal{T}'_l}\int_{}^{}\prod_{\substack{k=0\\k\neq a_l,b_l}}^{2X-1}dz_k~g^{(X)}_{\mathcal{B},\mathcal{T}'_l}(\mathcal{Z})~,
	\label{eq:G_phy}
\end{equation}
where
\begin{align}
	g^{(X)}_{\mathcal{B},\mathcal{T}'_l}(\mathcal{Z}) =& \,\, i^X (-1)^{N_\times(\mathcal{T})+N_<(\mathcal{B},\mathcal{T},\mathcal{Z})} \nonumber\\
	&\times \left[ \prod_{\substack{k=0\\k\neq l}}^{X-1} \Delta(z_{a_k},\alpha_{a_k}; z_{b_k},\alpha_{b_k}) \right] \nonumber\\
	&\times \Tr\left[\prod_{i=0}^{2X-1}\bm{\mathcal{G}}(z_{i+1}, z_{i}) \bm{F}^{\alpha_i}\right], 
	\label{eqn:underlyingGphy}
\end{align}
and $z_{2X}=z_0$~.

\subsection{\texorpdfstring{$X$}{X}OA diagram generation}
\begin{figure}[tb]
	\centering
	\includegraphics[width=0.45\textwidth]{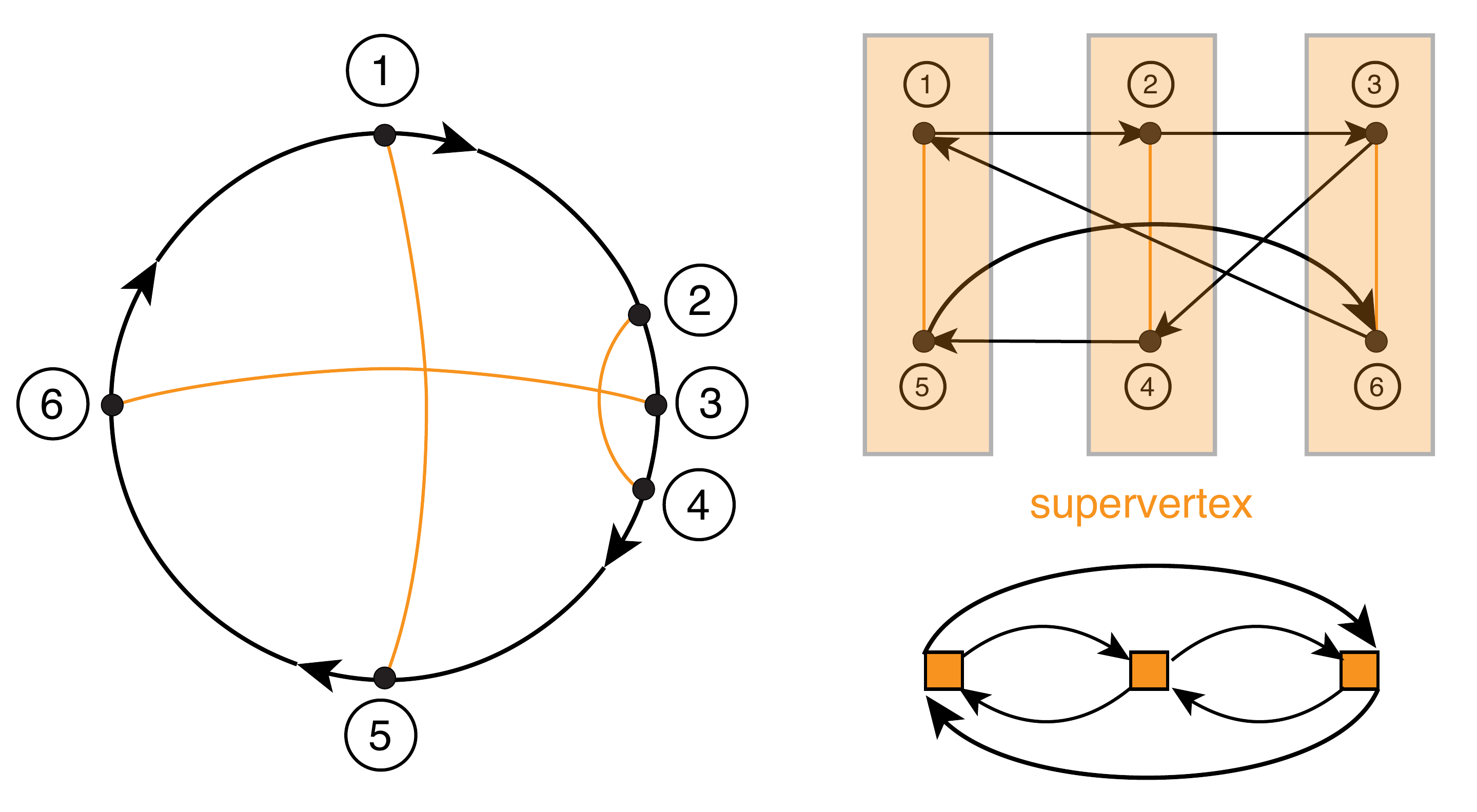}
	\caption{Transformation of a 3OA graph for checking the 2PI condition. The left panel shows the original diagram, the top right panel the definition of the supervertices (orange boxes), and the bottom right panel the corresponding supergraph consisting of supervertices connected by PP propagators. 
	}
	\label{fig:supervertex}
\end{figure}
In order to generate the PP self-energy diagrams of order $X$, we first generate all possible $(2X-1)!!$ diagram topologies $\mathcal{T}$ via the depth-first-search~\cite{Tarjan1972}, disregarding the 2PI condition.
For a given topology or set of hybridization connections, we construct a graph whose supervertices consist of pairs of vertices connected by hybridization lines, see orange boxes in Fig.~\ref{fig:supervertex}.
Now the corresponding adjacent matrix represents the connection between those supervertices by PP propagators. Since the $X$OA diagram is composed of dressed propagators, each of them includes self-energy corrections.
To avoid a double counting of self-energy contributions, we select the so-called two-particle irreducible (2PI) diagrams by investigating the three-edge-connectivity of the supergraph.
We use an $X$OA-adapted implementation~\cite{Kim2022} of the three-edge-connectivity algorithm~\cite{Tsin2007,Norouzi2014}, whose computational cost scales linearly as a function of the number of supervertices.

\begin{figure}[tb]
	\centering
	\includegraphics[width=0.45\textwidth]{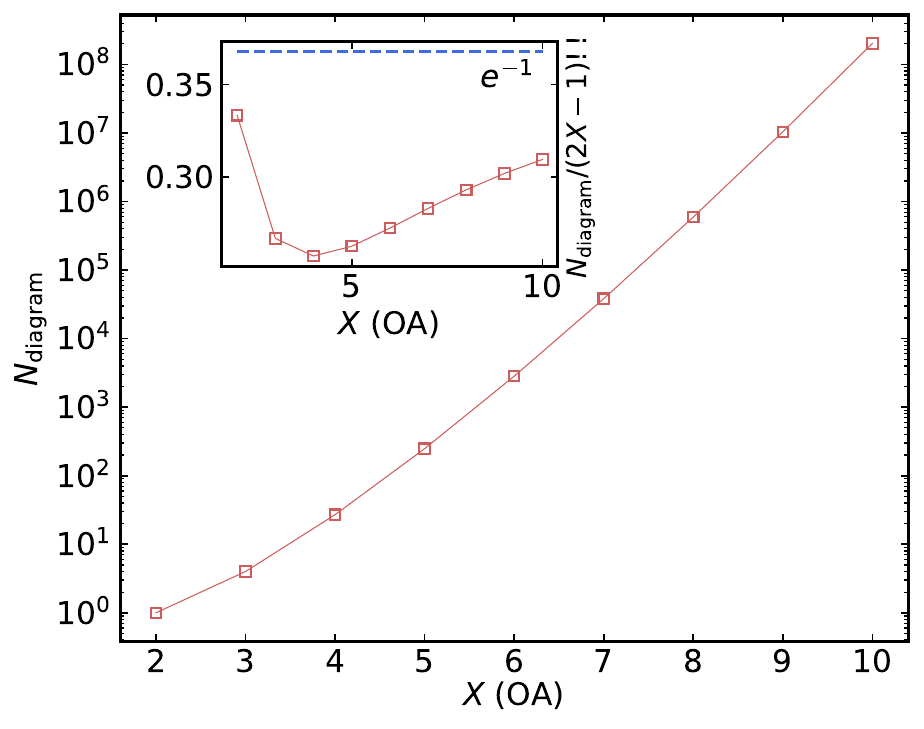}
	\caption{
		Number of $X$OA diagrams as a function of diagram order $X$.
		The inset shows the ratio between the number of 2PI diagrams and the total (including two-particle reducible) number of diagrams. The dashed line in the inset shows the exact asymptotic value of the ratio~\cite{Kleitman1970}.
	}
	\label{fig:Ndiagram}
\end{figure}

We confirmed that the number of irreducible diagrams generated by this algorithm, listed in Table~\ref{tab:Ndiagram}, matches the exact recurrence relation [Eq.~(2.1) in Ref.~\cite{Stein1978}] at least up to diagram order 10.
Although our order window is still too small to approach the exact asymptotic limit, $e^{-1}(2X-1)!!$ (see inset of Fig.~\ref{fig:Ndiagram})~\cite{Kleitman1970}, it is clear that the number of diagrams increases \textit{factorially} and that the ratio between the number of 2PI diagrams and the total number ($(2X-1)!!$) remains at the level of $25\%\sim35\%$. 

Computing a tensor train for each of the factorial number of 2PI diagrams would quickly surpass our computational resources.
In the current implementation, we therefore sum those $N_{\text{diagram}}$ diagrams and use this sum as the underlying function for the construction of a single tensor train.

\begin{table}[b]
	\centering
	\begin{tabular}{cc|cc}
		\hline\hline
		$X$ & $N_{\text{diagram}}$ & \hspace{5pt} $X$ & \hspace{2pt} $N_{\text{diagram}}$\\
		\hline
		1 & 1 &		\hspace{5pt} 6  & \hspace{2pt} 2830\\
		2 & 1 &		\hspace{5pt} 7  & \hspace{2pt} 38232\\
		3 & 4 &         \hspace{5pt} 8  & \hspace{2pt} 593859\\
		4 & 27 &        \hspace{5pt} 9  & \hspace{2pt} 10401712\\
		5 & 248 &       \hspace{5pt} 10 & \hspace{2pt} 202601898\\
		\hline\hline
	\end{tabular}
	\caption{The number of $X$OA diagrams as a function of $X$.
	}
	\label{tab:Ndiagram}
\end{table}

\subsection{Variable-separated quantics tensor-cross interpolation}
The integrand of the PP self-energy [Eq.~(\ref{eqn:underlyingSigma_pp})] or physical Green's function [Eq.~(\ref{eqn:underlyingGphy})] can be viewed as a $(2X-1)$-variable function of contour times after fixing the first contour time as the reference.
Upon discretization of the time contour into $2^N$ steps, one can further express each time variable in the binary representation, which leads to a multi-bit function to be summed.
We decompose this multi-bit function into a matrix product (or tensor train) form via quantics tensor-cross interpolation (QTCI)~\cite{Oseledets2010,Fernandez2022,Shinaoka2023}.

\begin{figure}[t]
	\centering
	\includegraphics[width=0.49\textwidth]{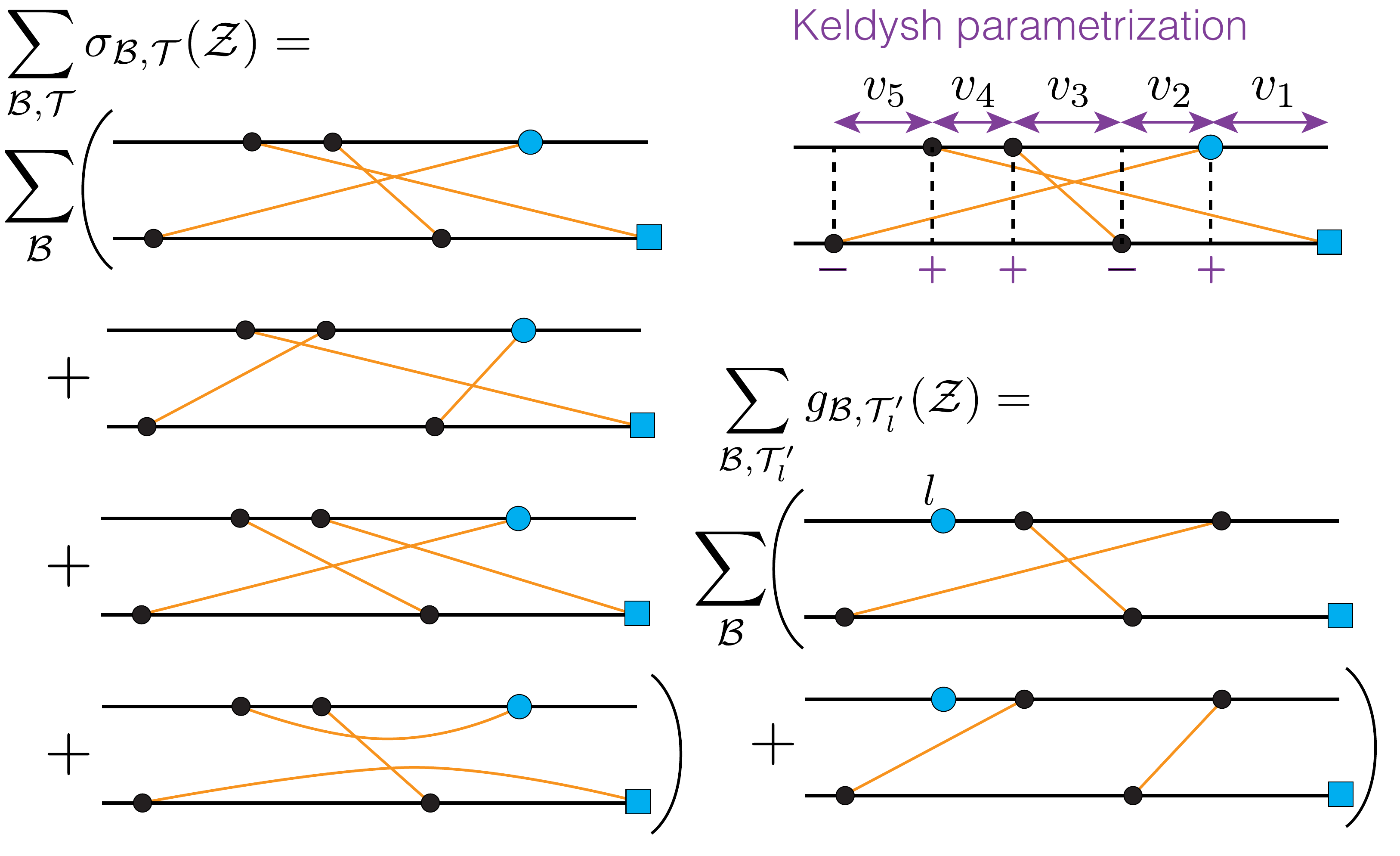}
	\caption{Diagram-summed contribution of a 3OA configuration to the PP self-energy (left) and Green's function (bottom right). The top right panel illustrates the reparametrization of the configuration in terms of Keldysh indices and physical times.
	}
	\label{fig:diagram_sum}
\end{figure}

The bond dimension of tensors in the tensor train, the key factor determining the efficiency of the QTCI based algorithm, strongly depends on the parametrization.
In the first QTCI version of the strong-coupling expansion up to 2OA~\cite{Kim2025}, the bits of the contour-ordered times~(Fig.~\ref{fig:diagram}) were arranged in the standard scale-separated sequence \cite{Shinaoka2023}.
In this representation, the greater and lesser components of the PP self-energy (or physical Green's functions) are encoded as separate domains in a single function. The resulting step-function like discontinuity amplifies the bond dimensions of the tensor-train representation. Switching to a Keldysh parameterization of the contour times (Fig.~\ref{fig:diagram_sum}) \cite{Eckstein2024} and performing separate fits for each combination of Keldysh indices resolved the domain problem, and significantly reduced the bond dimensions, which allowed to access the 3OA level \cite{Geng2025}.
On the other hand, Ref.~\cite{Eckstein2024} used the variable-separated scheme without quantics representation and solved a Holstein atom up to 4OA, at least for the greater component. 
In this work, we combine the QTCI and the variable-separated parametrization, which leads to a further reduction of the bond dimensions and enables 4OA calculations for generic Hubbard models in DMFT (5OA calculations in parameter regimes with low bond dimensions).

It turns out that there exists an ergodicity problem in the pivot search during QTCI that is particularly severe in the variable-separated scheme.
We tested two different approaches to overcome this ergodicity problem, first by introducing random noise whose amplitude is suppressed with increasing number of function calls, and second by introducing the random pivot search suggested in Ref.~\cite{Ishida2025}. Both methods resolve the ergodicity problem, but the latter approach is more advantageous since it does not compromise the fitting accuracy by the noise in the underlying function.

\begin{figure}[t]
    \centering
    \includegraphics[width=0.45\textwidth]{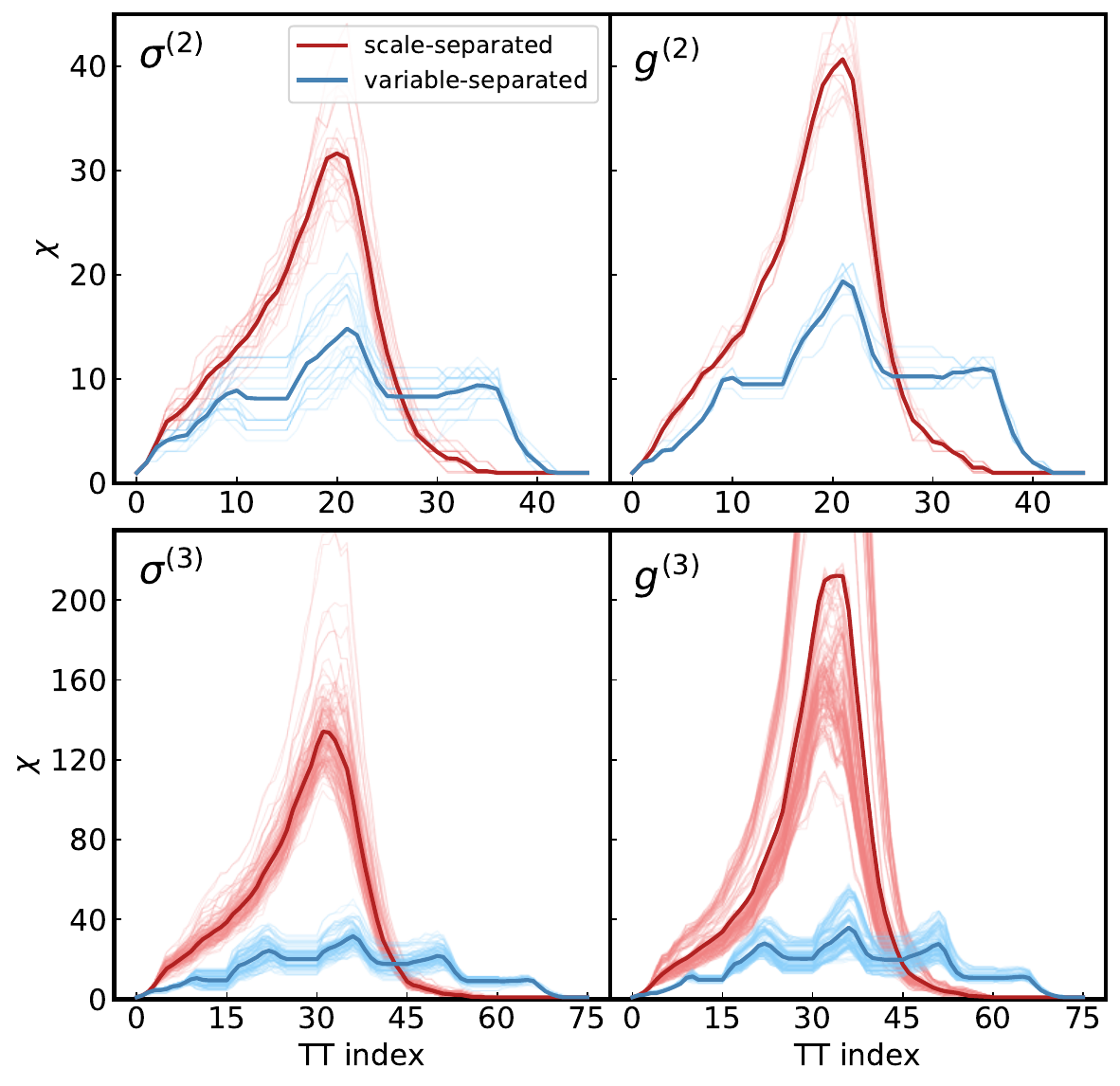}
    \caption{
	    Bond dimension $\chi$ required to represent $\sigma^{(X)}$ (left column) and $g^{(X)}$ (right column) of the half-filled Anderson impurity model using the scale-separated and variable-separated schemes for $X=2$ (top row) and $3$ (bottom row). The impurity site with $U=2$ and $\mu=1$ is coupled to the fermionic bath of the semi-circular density of states (the half-bandwidth $D$ as the energy unit) via the constant coupling $g=0.8$. The temperature is $T=0.1$.
	    Each tensor train was constructed via QTCI with a maximum pointwise error of $\sim 10^{-4}$, verified against $10^{6}$ randomly sampled tensor entries. Shaded lines show the bond-dimension profile of the individual tensor-trains; solid lines denote their average.
    }
    \label{fig:chiComparison}
\end{figure}

The variable-separated scheme is natural for the strong-coupling expansion, since the backbone part of the diagram is already factorized in this representation. 
Indeed, the variable-separated scheme shows much lower bond dimensions, compared to the scale-separated one, signaling a potentially easy decoupling of the 
hybridization part of the diagram weight~\cite{Kaye2024}.
Figure~\ref{fig:chiComparison} shows the bond dimensions $\chi$ of the converged PP self-energy and physical Green's function of the Anderson impurity model coupled to a bath with semi-circular density of states.
For a fair comparison, we fixed the absolute error bound of the underlying function, defined by the maximum absolute difference between the exact function and the tensor-train function fit, around $10^{-4}$~.
The largest bond dimension of the variable-separated scheme 
averaged over all individual tensor trains 
is about half of that of the scale-separated one at the 2OA level, and smaller by a factor of about $5$ ($\Sigma_{\text{pp}}$) to $8$ ($G_{\text{phy}}$) at the 3OA level.
Since the cost of QTCI fitting scales as $O(\chi^3)$~\cite{Fernandez2025}, such a reduction is expected to yield a substantial speed-up in computational time.
Actual
wall-clock times show a $\sim\!1.7$--$1.8\times$ speed-up for the variable-separated scheme at the 2OA level, and
a $\sim\!4\times$ ($\Sigma_{\text{pp}}^{(3)}$) to $\sim\!15\times$ ($G_{\text{phy}}^{(3)}$) speed-up at the 3OA level.
Importantly, the advantage of the variable-separated scheme grows markedly with increasing diagram order, both in bond dimension and in computational cost.
We did not 
compare the bond dimensions beyond 3OA since the calculation time for the scale-separated scheme grows out of reach.

It is also noteworthy that the variation of the bond dimension with the tensor train index is clearly different for the variable- and scale-separated schemes.
While there exists a single peak in the scale-separated scheme, the variable-separated scheme exhibits $2X-1$ peaks separated by valleys corresponding to the switching points between neighboring contour-time variables.

\subsection{Diagram-summed parametrization}

In Eq.~(\ref{eq:sigma_pp}) and (\ref{eq:G_phy}), each topology $\mathcal{T}$ of the PP self-energy ($\sum^{}_{\mathcal{B}}\sigma_{\mathcal{B},\mathcal{T}}(\mathcal{Z})$) or $\mathcal{T}'_l$ of the physical Green's function ($\sum^{}_{\mathcal{B}}g_{\mathcal{B},\mathcal{T}'_l}(\mathcal{Z})$) can be represented by a single tensor train, which we call the \textit{diagram-separated} scheme, 
or all topologies can be summed within a single tensor train,~($\sum^{}_{\mathcal{B},\mathcal{T}}\sigma_{\mathcal{B},\mathcal{T}}(\mathcal{Z})$ or $\sum^{}_{\mathcal{B},\mathcal{T}'_l}g_{\mathcal{B},\mathcal{T}'_l}(\mathcal{Z})$), which corresponds to the \textit{diagram-summed} scheme (Fig.~\ref{fig:diagram_sum}).
After integrating over the contour times $\mathcal{Z}$ (and summing over $\mathcal{T}$ or $\mathcal{T}'_l$ for the diagram-separated scheme), the resulting expressions are equivalent in both cases, but their computational cost differs, especially for higher orders. 

Each function call for both the PP self-energy and the physical Green's function contains two dominant parts: the backbone and topology (hybridization function) parts.
In the diagram-summed QTCI, one can recycle the backbone calculation, since some backbone configurations can be shared by different topologies, while in the diagram-separated approach, the same backbone calculation is repeated $N_{\text{diagram}}$ times for the different topologies.
The recycled backbone calculation provides a computational advantage to the diagram-summed QTCI, which grows with increasing diagram order $X$.
The first column of Table~\ref{tab:speedup} (ftn. call) presents the relative speed-up of the diagram-summed function call compared to the diagram-separated one, defined as
\begin{align}
	&\text{ftn. call speed up} = 
	\left\{
	\begin{array}{ll}
		\frac{N_{\text{diagram}}\times t_{\text{sep}}}{t_{\text{sum}}} & (\text{for } \Sigma_{\text{pp}})\\[0.5em]
		\frac{N_{\text{diagram}}\times t_{\text{sep}}}{N_{\text{ring}}\times t_{\text{sum}}} & (\text{for } G_{\text{phy}})
	\end{array}
\right.~,
	\label{eq:ftn_call_speedup}
\end{align}
where $t_{\text{sum}}$ ($t_{\text{sep}}$) is the time for the diagram-summed (diagram-separated) function call, averaged of the all tensor trains. 
One can see the rapidly increasing computational gain particularly for the PP self-energy.
The significantly smaller computational gain for the physical Green's function can be attributed to the smaller number of summable diagrams;
for a given backbone configuration, only the subset of diagrams that share the same position of the external (amputated) vertices can be summed, see Fig.~\ref{fig:diagram_sum} for 3OA, and there are $N_{\text{ring}}=2X-3$ such distinct positions.

\begin{table*}[hbtp]
    \centering
    \begin{tabular}{c|ccc|ccc|ccc}
        \hline\hline
	\multirow{2}{*}{speed up} & \multicolumn{3}{c|}{3OA} & \multicolumn{3}{c|}{4OA} & \multicolumn{3}{c}{5OA}\\
	& ftn call & $\chi^3$ ratio & total & ftn call & $\chi^3$ ratio & total & ftn call & $\chi^3$ ratio & total \\
        \hline
	$\Sigma_{\text{pp}}$ & $3.2$ & 0.39 & 1.25 & $7.5$ & 0.21 & 1.58 & $11.2$ & - & -\\
        \hline
        $G_{\text{phy}}$ & $1.1$ & 0.64 & 0.70 & $2.5$ & 0.26 & 0.65 & $5.3$ & - & -\\
        \hline\hline
    \end{tabular}
    \caption{
	    Relative speed-up of the diagram-summed function calls for $\Sigma_{\text{pp}}^{(X)}$ and $G_{\text{phy}}^{(X)}$, compared to the diagram-separated ones at the 3OA/4OA/5OA level.
	    The function call speed up is defined in Eq.~(\ref{eq:ftn_call_speedup}).
    The $\chi^3$ ratio compares the cubic power of the largest (averaged) bond dimensions of the two schemes for the Anderson impurity model data shown in Fig.~\ref{fig:DiagramSumComparison}.
    }
    \label{tab:speedup}
\end{table*}

\begin{figure}[b]
    \centering
    \includegraphics[width=0.45\textwidth]{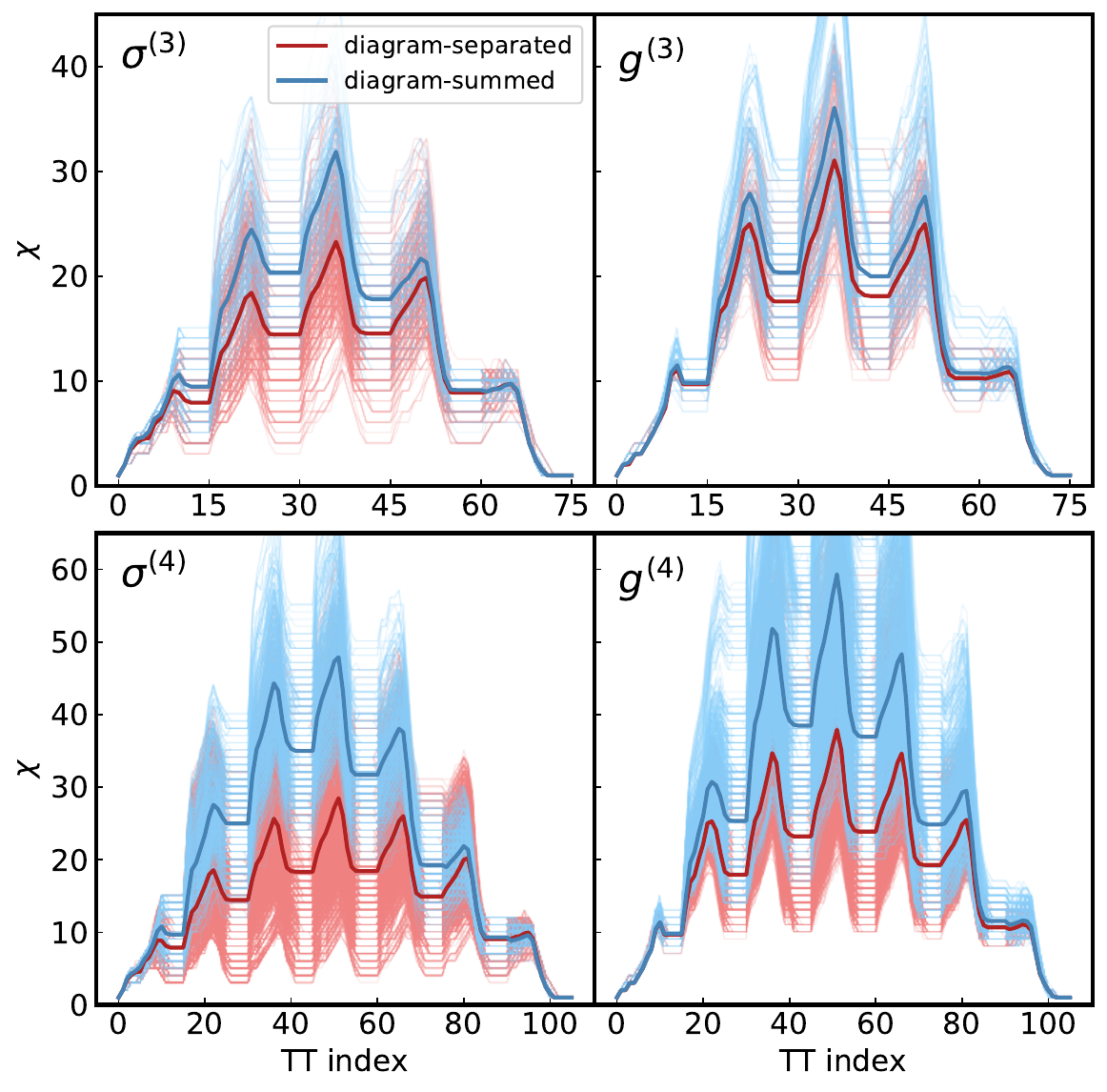}
    \caption{
	    Comparison of the bond dimension $\chi$ for $\sigma^{(X)}$ (left column) and $g^{(X)}$ (right column) in the diagram-separated and diagram-summed representations, for $X=3$ (top row) and $X=4$ (bottom row).
	    We use the Anderson impurity model with the same parameter set as in Fig.~\ref{fig:chiComparison}.
    Both representations use the variable-separated scheme.
    Shaded lines show the bond dimensions of the individual tensor trains and the solid lines show their average.
    }
    \label{fig:DiagramSumComparison}
\end{figure}

However, this computational gain can be compensated by the increasing bond dimension of the tensor trains.
Figure~\ref{fig:DiagramSumComparison} shows the bond dimensions of the quantics tensor trains for the Anderson impurity model.
We find that summing the diagram topologies typically leads to an increase of the bond dimension both for the PP self-energy and the physical Green's function.
Since both the QTCI fitting and subsequent integration have a computational cost $O(\chi^3)$ \cite{Fernandez2025}, the second column of Table~\ref{tab:speedup} presents the computational loss due to the increased bond dimension.
Considering both the number of function calls and the effect on the bond dimension, we conclude that the diagram-summed (diagram-separated) scheme is beneficial for the PP self-energy (physical Green's function).

\section{Results}\label{sec:results}
\subsection{\texorpdfstring{$U=0$}{U=0} limit}
As a nontrivial benchmark for the strong-coupling expansion, we investigate the $U=0$ limit of the paramagnetic DMFT solution on the Bethe lattice.
We specifically aim to recover the exact spectral function by extrapolating the diagrammatic series starting from the exact PP Green's function $\mathcal{G}^*$ and the exact hybridization function $\Delta^*$.
For this task, we introduce the expansion parameter $\xi$ in front of the hybridization function, $\Delta(\omega)\rightarrow \xi\Delta(\omega)$, express the spectral function for a given frequency as a power series of $\xi$
\begin{equation}
	A(\omega;\mathcal{G}^*,\Delta^*) = \sum^{k_{\text{max}}}_{k=0}a_k(\omega;\mathcal{G}^*,\Delta^*)\xi^k~,
	\label{eq:xi_power_series}
\end{equation}
and extrapolate to the $k_{\text{max}}\rightarrow\infty$ limit at $\xi=1$ using the well-established Pad\'e-based method~\cite{Baker1961,Hunter1973,Hunter1979}.
In general, the exact PP Green's function and hybridization function are unknown a priori.
For each self-consistent iteration, the diagrammatic series could be extrapolated to infinite order, after which the $\mathcal{G}$ and $\Delta$ are updated.
The $U=0$ case is a limiting case where we know the exact $\Delta$, and the procedure can hence be simplified.
The DMFT self-consistency relation for the Bethe lattice
\begin{equation}
	\Delta(\omega) = \left(\frac{D}{2}\right)^2G_{\text{phy}}(\omega)~,
	\label{eq:bethe_selfcons}
\end{equation}
which follows from the semi-circular density of states of the noninteracting lattice~\cite{Georges1996} (the half-bandwidth $D$ being the energy unit throughout the paper),
\begin{equation}
	A(\omega) = -\frac{1}{\pi}\mathrm{Im} G^R_{\text{phy}}(\omega) = \frac{2}{\pi D}\sqrt{1-\left(\frac{\omega}{D}\right)^2},
	\label{eq:semicircle}
\end{equation}
allows to compute the exact hybridization function.
Instead of the exact PP Green's function, we use the 5OA PP result, which should be close enough to the exact one to enable a meaningful analysis and extrapolation of the series. 

\begin{figure}[t]
	\centering
	\includegraphics[width=0.48\textwidth]{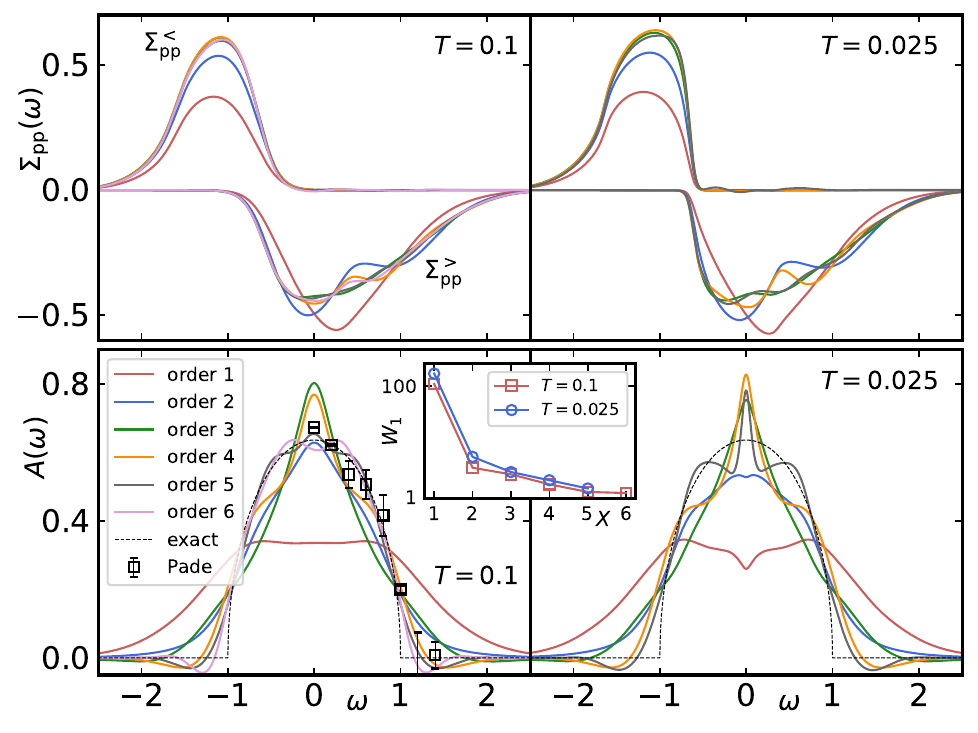}
	\caption{
		Noninteracting ($U=0$) PP self-energy (top) and spectral function (bottom) for $T=0.1$ (left) and $0.025$ (right).
		Since all PP are equivalent for the half-filled $U=0$ case, we suppress the PP index.
		(Black) squares show the infinite-order extrapolated value. The error bars are defined by the maximum and minimum of two diagonal Pad\'e orders: [3/2] and [2/3].
		The inset of the lower panels shows the Wasserstein metric $W_1$ between the spectral function at the given order and the exact semi-circular one.
	}
	\label{fig:u_zero_limit}
\end{figure}

In the top left panel of Fig.~\ref{fig:u_zero_limit}, one can see that for temperature $T=0.1$, the 5OA provides an almost converged solution for the self-energy. 
The convergence of the spectral function (lower left panel) is slower and oscillating, but the spectra seem to systematically converge to the exact reference (dashed line). 
As a measure quantifying the spectral distance from the exact one, we use the Wasserstein metric ($W_1$) that compares the absolute difference between two cumulative distribution functions~\cite{Kantorovich2006,Villani2009}.
The inset of the lower panels shows the systematic decrease in $W_1$ between the exact semi-circular density of states and the spectral function $A(\omega)$ as a function of the diagram order.
At the lower temperature, $T=1/40$ (right panels in Fig.~\ref{fig:u_zero_limit}), even higher orders are needed to converge the self-energy and spectral function. However, $W_1$ decreases toward 0 with increasing order $X$, and there 
is no clear evidence of misleading convergence of the strong-coupling series~\cite{Schafer2013,Kozik2015}. 

The lower left panel of 
Fig.~\ref{fig:u_zero_limit} shows that the strong-coupling diagrammatic series starting from the exact hybridization and 5OA PP Green's function approaches  the exact semi-circular density of states.
Each line of this panel represents the partial sum of the spectral function from order 1 to 6, sharing the same input.
The extrapolated values, with error bars extracted from two different Pad\'e fits, are consistent with the exact spectral function, except at $\omega=0$. At the lower temperature (lower right panel), a reliable extrapolation is not yet possible, but the evolution of the spectra with increasing order again suggest a systematic convergence towards the exact $U=0$ result.  
We note that in our previous study up to the third-order approximation (3OA)~\cite{Geng2025}, it was unclear if the strong-coupling expansion can recover the exact solution at $U=0$, due to a lack of sufficiently high diagram orders.

\subsection{Bond dimensions of DMFT solutions}\label{sec:bondDim}
\begin{figure*}[tb]
	\centering
	\includegraphics[width=0.6\textwidth]{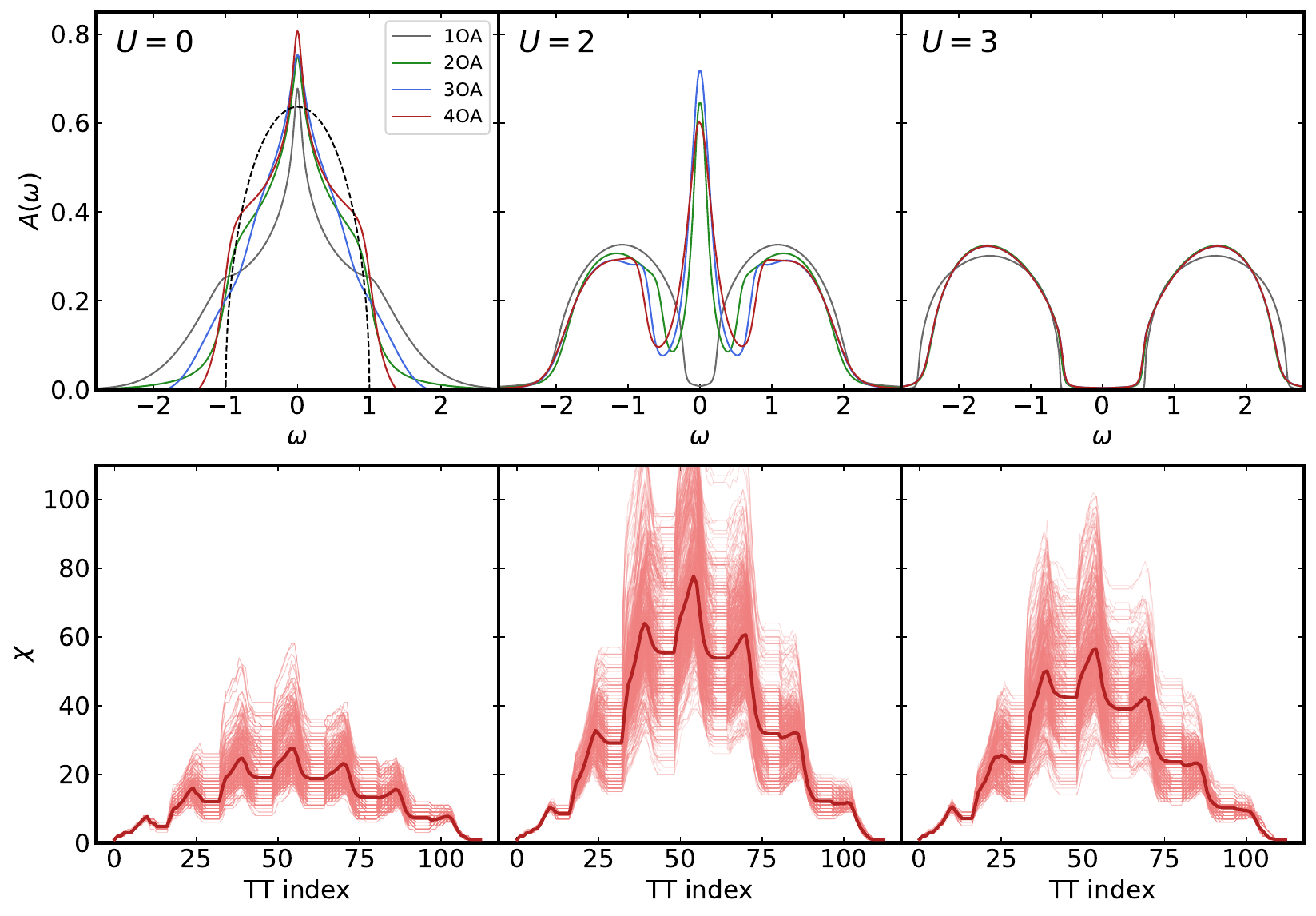}
	\caption{
        (Top) Spectral functions of three representative paramagnetic phases of the infinite-dimensional Hubbard model on the Bethe lattice for $T=0.025$: noninteracting metal, correlated metal and Mott insulator, computed self-consistently at successive expansion orders (1OA--4OA, see legend). For the noninteracting ($U=0$) case, we use the exact hybridization function [Eq.~\eqref{eq:bethe_selfcons}], since the self-consistently determined one leads to convergence problems. The black dashed line in the left panel is the known exact spectral function [Eq.~\eqref{eq:semicircle}], the same exact reference used as the benchmark target in Fig.~\ref{fig:u_zero_limit}. (Bottom) Corresponding bond dimensions $\chi$ of the fourth-order contribution $g^{(4)}$ to the converged 4OA solution.
		Thin lines show the bond dimensions of the individual tensor trains and the thick lines show their average.
	}
	\label{fig:variousBondDim}
\end{figure*}

The convergence rate of the strong-coupling expansion is not directly correlated with the required bond dimension for the tensor-train representation.
Figure~\ref{fig:variousBondDim} compares the bond dimension for the physical Green's function and the convergence behavior of three representative DMFT solutions of the paramagnetic Hubbard model on the Bethe lattice: a noninteracting metal, a correlated metal, and a Mott insulator.
Since the zeroth-order reference system of the strong-coupling expansion is the isolated impurity, the Mott-insulating solution exhibits rapid convergence with increasing expansion order $X$. 
Even 2OA is enough to capture the detailed spectral structure near the band edges.
In the correlated-metal and noninteracting regimes, however, the system becomes more itinerant and high-order corrections are required for quantitative accuracy.
For $U=2$, although the functional form (a central quasi-particle peak coexisting with upper and lower Hubbard bands) is settled since 2OA, the peak heights and positions continue to change at the 4OA level.
In the noninteracting case, the spectral function is still far from the exact one even for 4OA.
(Note that in contrast to Eq.~\eqref{eq:semicircle}, we here show the self-consistently computed $X$-order approximations, not the solutions for the exact hybridization function and 5OA PP propagators. But we still use the exact hybridization function for the $U=0$ case, where the self-consistently determined hybridization function leads to convergence problems.)

Interestingly, the required bond dimension does not change monotonically as $U$ decreases.
It first increases from the Mott insulator to the correlated metal, then decreases sharply toward the noninteracting limit, where the smallest bond dimensions are observed.
This shows that slow convergence of the strong-coupling series does not necessarily imply a high-rank tensor representation.
The maximum bond dimension rather correlates with the complexity of the correction to the 
spectral function (and hence Green's function or hybridization function) at the highest order, e.g. the fourth-order contribution $A^{(4)}$.
At least, it increases with the number of sharp and broad peaks appearing in $A^{(4)}(\omega)$ (not shown).

\subsection{Photo-doped Mott insulator}
We also consider photo-doped Mott insulators, where the non-equilibrium steady state is achieved by fixing a double-step Fermi distribution function~\cite{Erpenbeck2023}.
By varying the doublon and holon Fermi levels located at $\pm \mu_{\text{ph}}$, we can effectively control the doublon and holon density.
In previous studies, it has been shown that clear quasi-particle peaks develop near these Fermi levels. However, the character of the quasi-particle changes depending on the photo-doping level; for low photo-doping, the introduced doublons and holons play the role of mobile charge carriers, while the quasi-particle peak is associated with the singlons at high photodoping.

\begin{figure*}[t]
	\centering
	\includegraphics[width=0.60\textwidth]{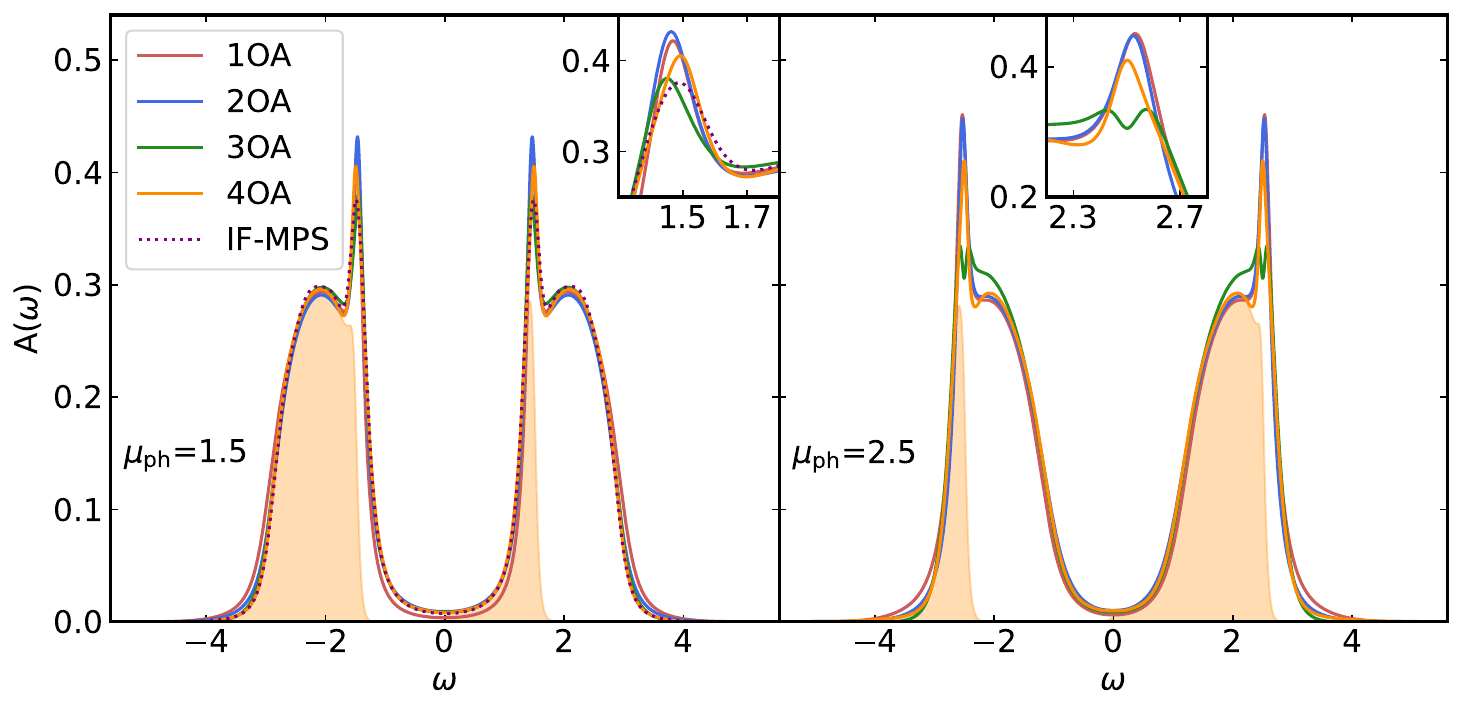}
	\caption{
		Spectral functions of the photo-doped Mott insulator for $T=0.05$ and $U=4$. 
		The left and right panels show representative results for low- and high-photodoping cases with $\mu_{\text{ph}}=1.5$ and $2.5$, respectively.
		In the left panel, the purple dashed line presents the spectral function obtained with the influence functional approach, taken from Ref.~\cite{Nayak2025}.
	}
	\label{fig:photodoping}
\end{figure*}
Figure~\ref{fig:photodoping} shows the spectral function of the photo-doped Mott insulator for representative low ($\mu_{\text{ph}}=1.5$) and high ($\mu_{\text{ph}}=2.5$) doping levels.
Although the effects of the doublon/holon doping to the singly occupied Mott insulator and singlon doping to the fully photodoped system are supposed to be equivalent in the $U\rightarrow \infty$ limit, the effects of high-order corrections turn out to be significantly inequivalent for finite $U$, even within the Mott insulator.
In the low-photodoping case, high-order corrections beyond 1OA do not produce qualitative changes; the quasi-particle peak height and position change only slightly and in an oscillatory manner.
The peak position of the 4OA becomes close to the IF result~\cite{Nayak2025}, while the agreement in the peak height is less conclusive, partly because
the IF spectrum was obtained from a real-time simulation on a much shorter time interval ($t_{\text{max}}\sim 60$) than in the present study ($t_{\text{max}}\sim 800$).

On the other hand, in the strong photo-doping case, the 3OA contribution drastically suppresses the quasi-particle peak and even results in a pseudogap \cite{Geng2026}.
However, the 4OA contribution restores the quasi-particle peak at a frequency which is slightly shifted to lower energy, in contrast to the shift toward higher energy in the low-photodoping case. 
We expect that with further increasing orders, the peak height will slowly converge toward an intermediate value in an oscillating manner.

\subsection{Photo-doped AFM insulator}
\begin{figure}[t]
	\centering
	\includegraphics[width=0.48\textwidth]{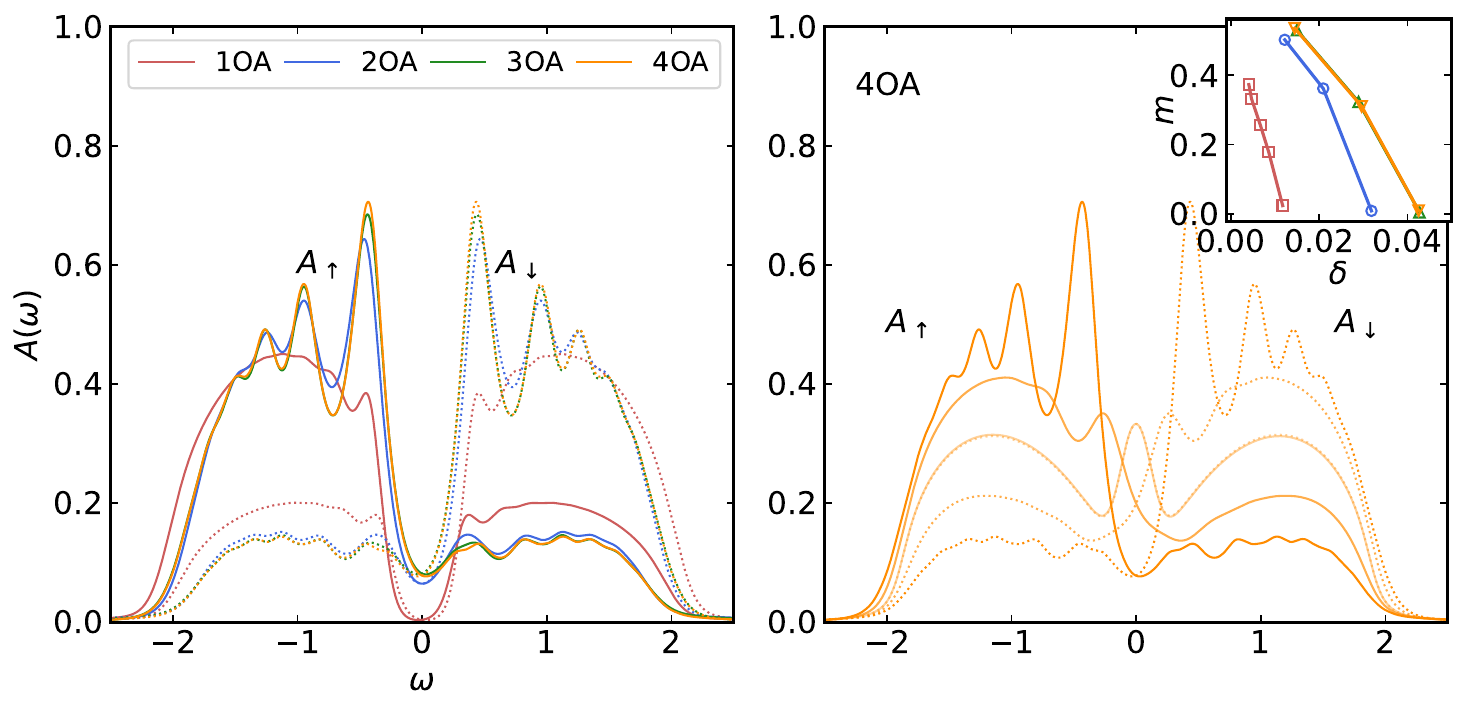}
	\caption{
		Spectral function and magnetization for $T=1/11$ and $U=2$.
		The left panel presents the $X$OA ($X=1-4$) spectra without photodoping and the right panel shows the photodoping dependence of the 4OA spectrum.
		In the right panel, the three solid (dashed) lines denote the spin-up (-down) equilibrium spectral function ($\delta=0.015$) and the photo-doped ones with 
		$\delta=0.03$ and $0.043$, respectively.
		The inset shows the continuous transition of the AFM order as a function of the photodoping concentration.
	}
	\label{fig:afm_photodoping}
\end{figure}
As a challenging application of the $X$OA-QTCI scheme, we investigate the photo-doped antiferromagnetically (AFM) ordered state.
Figure~\ref{fig:variousBondDim} shows that a complex spectral function with sharp peaks leads to a significant increase in the bond dimension.
The long-lived spin-polaron modes in the AFM-ordered background, whose energy spacing is proportional to exchange coupling scale $J_{\text{ex}}\sim t^2/U$, renders the spectral function of the antiferromagnetic system particularly complex, as shown in Fig.~\ref{fig:afm_photodoping}. Furthermore, it has been reported that a small causality breaking in the PP self-energy, e.g. due to an insufficient time grid, can spuriously destroy the AFM order~\cite{Geng2025}.

In this work, we employ the matrix-product operator (MPO) of the Fourier transform, which transforms the tensor-train with time variables into the one with frequencies, incorporating the trapezoidal quadrature rule.
It turns out that this MPO has a low bond dimension ($\sim 12$), if one pairs the largest scale bit of the time variable with the smallest scale bit of the frequency variable~\cite{Shinaoka2023}.
With this trick it becomes possible to calculate the AFM solution at the 4OA level.

The left panel of Fig.~\ref{fig:afm_photodoping} presents the spectra of the $X$OA AFM-ordered solutions in the intermediate-correlation regime.
As $X$ increases, the spin-polaron peaks become more prominent and the 4OA solution appears to be almost converged.
Compared to the 1OA result, it turns out that the high-order corrections enhance the magnetization for this parameter set.
In fact, the exact antiferromagnetic transition temperature reaches its highest value near $U=2$, while the overestimation of correlations in the 1OA leads to a shift of the maximum to lower $U$ \cite{Werner2012,Geng2025}. Thus, for fixed $U=2$ and $T=1/11$, 
the system shifts deeper into the AFM phase with high-order corrections.

The doublons and holons injected by the photo-doping suppress the AFM order, since their motion disturbs the spin background. 
As a function of the effective chemical potential, the magnetization monotonically decreases and the system shows a continuous transition to a photo-induced paramagnetic state. 
Compared to the 1OA, the magnetization and its critical photodoping concentration $\delta$, defined by the electron occupation for $\omega>0$, is larger in the 4OA, consistent with the above observation of a shift deeper into the AFM phase.

\section{Conclusion}
\label{sec:conclusions}
We presented a real-time strong-coupling impurity solver for general steady states including equilibrium and photo-doped states, as well as long-range ordered (AFM) phases.
The variable-separated QTCI scheme supplemented by the random pivot search drastically reduces the maximum bond dimension, at higher orders by more than a factor of five. This corresponds to a reduction of the computational cost by about a factor of $5^3$, allowing us to include the previously inaccessible fourth-order diagrams, and for one-shot calculations the diagrams up to sixth order. 
We also implemented a depth-first-search-based algorithm for the automated high-order diagram generation, and proposed the diagram-summed scheme as a strategy to mitigate the factorially increasing number of diagrams at higher orders.

With these technical improvements, the accessible diagram orders are now comparable to the current state-of-the-art inchworm algorithm (up to the fifth order for the PP self-energy and the third for the physical Green's function)~\cite{Kunzel2024}. It will be interesting to systematically benchmark the two methods, but an advantage of the QTCI algorithm is that it is free from stochastic Monte Carlo errors, while the fitting errors in the multi-dimensional integrands to some extent average out in the calculation of the integrals.
 
Using the new algorithm, we demonstrated the systematic convergence in the noninteracting limit, where the recovery of the exact solution via the self-consistent strong-coupling expansion has remained an open question, and also in the case of photo-doped Mott insulators. Especially in the case of large photodoping, it turns out to be important to include high-order diagrams beyond the 3OA, since the convergence is slow and oscillatory.
As a final example, we studied the effect of photodoping on the AFM-ordered state. Consistent with previous studies at the 1OA level, we found that the excess doublons and holons introduced by photodoping systematically reduce the magnetic order, and completely suppress it beyond a critical photo-doping concentration. Compared to the 1OA (NCA) results, the magnetization in the intermediate-coupling regime is increased and the spectral function shows more prominent spin-polaron features. Accurately describing the equilibrium and photo-doped AFM state has not only been a challenge for the self-consistent strong-coupling expansion, but also for other state-of-the-art real-time impurity solvers, such as the recently developed influence functional approach \cite{Thoenniss2023,Nayak2025}. 

As we have shown in the noninteracting case, the strong-coupling expansion now reaches a level where an extrapolation of the diagrammatic series to infinite order becomes feasible.
Although (diagrammatic) series extrapolations have been exploited in the context of various numerical methods, including diagrammatic Monte Carlo~\cite{Prokofev1998,VanHoucke:2010ky,Kozik:2010fla}, the high-temperature expansion~\cite{Henderson1992}, and the numerical linked-cluster method~\cite{Rigol2006,Rigol2007}, developing a stable extrapolation scheme for the continuous spectral function is an open problem. Finding a robust solution to this problem is especially important to self-consistent iteration schemes like $X$OA.

\section{Acknowledgments}
This work was supported by Basic Science Research Program through the National Research
Foundation of Korea (NRF) funded by the Ministry of Education (2025090055,2026030151), and the Swiss National Science Foundation via NCCR Marvel and Grant
No. 2000-1-240023. The calculations were run on the bear and iREMB
clusters at the DGIST.

\bibliography{ref}
\end{document}